\documentclass{PoS}

\title{Simulation of lattice gauge action from the overlap operator}

\ShortTitle{Simulation of lattice gauge action from the overlap
operator}

\author{\speaker{K.F. Liu}\\
        Dept.\ of Physics and Astronomy, University of Kentucky,
        Lexington, KY 40506, USA\\
        E-mail: \email{liu@pa.uky.edu}}

\abstract{We show the result of the lattice gauge field tensor
which is derived from the classical continuum limit of the overlap
Dirac operator. By analogous construction, it was recently
proposed that the gauge action can be obtained from the overlap
operator as well which is proportional to ${\rm Tr}\, a D_{ov}$.
We discuss how to carry out Monte Carlo simulations of such a
gauge action together with the overlap fermion.}

\FullConference{XXIVth International Symposium on Lattice Field Theory\\
                July 23-28, 2006\\
                Tucson, Arizona, USA}

\begin{document}

\section{Introduction}

Lattice gauge actions are usually constructed from the link
variable $U$. Similarly, the gauge field tensor is constructed
from the clover-leaf like plaquettes.

The advent of Neuberger's overlap operator~\cite{neu98} has the
implication in a new direction. The overlap Dirac operator is
development to solve the fermion chirality problem in lattice
QCD~\cite{neu01}. It does not have $O(a)$ error for the fermion
operators. The small $O(m^2 a^2)$ errors and non-perturbative
renormalization via current algebra relations and Ward identities
make it a desirable fermion formulation for both light and heavy
quarks~\cite{liu05}. An index theorem was formulated by
Hasenfratz, Laliena, and Niedermayer~\cite{hln98} based on the
operator at finite cut-off. It is then pointed out by L\"{u}scher
that the anomalous behavior of the fermion partition function
under a flavor-singlet transformation is expressed by the index of
the the overlap Dirac operator arising from the Jacobian,
providing a clear understanding of the exact index
theorem~\cite{lus98,hln98} in the path-integral formalism.
Following developments have seen the explicit derivation of the
local lattice topological charge in terms of the local overlap
operator via weak coupling expansion by Kikukawa and
Yamada~\cite{ky99}, explicit calculations without gauge coupling
expansion by Adams~\cite{ada02}, Fujikawa~\cite{fuj99} and
Suzuki~\cite{suz99}, i.e.
\begin{equation}  \label{top}
q_L(x) = {\rm tr}_{cs}\, \gamma_5(1 - 1/2 a D_{ov}(x,x))
\end{equation}
where $D_{ov}$ is the overlap operator and the trace is over the
color and Dirac spin indices. The lattice topological charge
operator $q_L(x)$ so defined approaches the topological charge
density $q(x)$ at the continuum limit,
\begin{equation}
q_L(x)_{\stackrel{-\!\!\longrightarrow}{a \rightarrow 0}}
 a^4 q(x) + O(a^6).
\end{equation}
    This formulation of the topological charge operator has not only
been used to check the Atiya-Singer index theorem at finite
lattice cut-off~\cite{zbb02}, the topological
susceptibility~\cite{dgp05}, but has also been adopted to study
the local topological structure of the
vacuum~\cite{hdd02,hdd03a,hdd03b,hov05,haz05a,haz05b,ahz05}. It is
with this operator that the sub-dimensional coherent sign
structures have been discovered in 4-D
QCD~\cite{hdd03b,hov05,haz05a,haz05b} and 2-D CP(N-1)
model~\cite{alt05}. It is argued that the chiral filtering of the
overlap fermion is responsible for optimally filtering out the
ultraviolet fluctuation to allow the structures to be
revealed~\cite{hor06a}. Indeed, other conventional topological
charge operator constructed from the gauge links were used, but
could not decipher the curvilinear structure observed with the
overlap operator~\cite{alt05}. This leads to the possibility that
gauge field tensor and gauge action derived from the overlap
operator might be good alternatives to those from the gauge links
directly. Recently, I. Horv\'{a}th~\cite{hor06b} proposed a
formulation of lattice QCD wherein all elements of the theory
(gauge action, fermionic action, theta-term, and other operators)
are constructed from a single object, namely the lattice Dirac
operator $D_{ov}$.  In this talk, I will present the results of
the derivation of the gauge field tensor as the classical
continuum limit from the overlap operator. The detailed
derivation~\cite{lah06} and the numerical evaluation~\cite{ahl06}
are under preparation and will be posted on the arXiv soon. I will
also discuss how to simulate such an action with the Hybrid Monte
Carlo algorithm.

\section{Gauge Field Tensor}

      To begin with, we note that a gauge covariant operator
which is a functional of $U$ satisfies the condition
\begin{equation} \label{cov}
O(g^{-1}Ug) = g^{-1}O(U)g,
\end{equation}
where $g$ is the local gauge transformation. It is obvious that
the local operator with the color trace ${\rm tr}_c\, O(U)(x,x)$
is gauge invariant
\begin{equation} \label{inv}
{\rm tr}_c\, O(g^{-1}Ug)(x,x) = {\rm tr}_c\,
g^{-1}(x)O(U)(x,x)g(x) = {\rm tr}_c\, O(U)(x,x).
\end{equation}
Since the overlap operator, being a Dirac fermion operator, is
gauge covariant and is not ultra-local, it is expected that the
classical continuum limit of the trace in both the color and spin
indices ${\rm tr}_{cs}\, \Gamma D_{ov}(x,x)$ will be the lowest
dimensional local gauge operator which reflects the Lorentz
structure of the gamma matrix $\Gamma$. Thus, it is not surprising
that ${\rm tr}_{cs}\, \gamma_5 D_{ov}(x,x)$ gives the local
topological charge density at the continuum limit. Therefore, one
expects that~\cite{hor06b}
\begin{equation}   \label{action}
{\rm tr_{cs}}\, (D_{ov}(x,x)-D_{ov}^0(x,x))
_{\,\,\stackrel{\longrightarrow}{a \rightarrow 0}\,\,} a^4 tr_c
F_{\mu\nu}^2 + O(a^6),
\end{equation}
where $D_{ov}^0$ is the non-interacting overlap operator with
gauge link $U$ set to unity. This has been verified with constant
field and with numerical evaluation~\cite{ahl06}.

In addition to gauge invariant operators, one can obtain gauge
covariant operators as well. Since $D_{ov}(x,x)$ is gauge
covariant and a Lorentz scalar, one expects that
$tr_s\,\sigma_{\mu\nu}D_{ov}(x,x)$ with the spin index traced over
will result in a lowest dimensional gauge covariant operator with
the $\mu\nu$ indices in the classical continuum limit which is
just the gauge field tensor~\cite{hor06b}. In other words,
\begin{equation}
{\rm tr_{s}}\,
\sigma_{\mu\nu}D_{ov}(x,x)_{\,\,\stackrel{\longrightarrow}{a
\rightarrow 0}\,\,} a^2 F_{\mu\nu} + O(a^4).
\end{equation}
We should note that the possibility of obtaining the lattice gauge
field tensor from the overlap operator by expanding in $a$ was
first pointed out by Niedermayer~\cite{nie99}. Also, it was
suggested by Gattringer~\cite{gat02} that the field tensor can be
defined through the square of various lattice Dirac operators,
i.e. $tr_s\, \gamma_{\mu}\gamma_{\nu}D^2(x,y)$ with a weighted sum
over $y$.

We have derived the classical limit following Adams~\cite{ada02}
and Suzuki's method~\cite{suz99}. While details of the derivation
will be given elsewhere~\cite{lah06}, we shall present the results
here.

\begin{eqnarray}
&& {\rm tr_{s}}  \,\sigma_{\mu\nu}\, a D_{ov}(x,x)
_{\stackrel{-\!\!\longrightarrow}{a \rightarrow 0}} \nonumber \\
&&\int_{-\pi}^{\pi}\frac{d^4k}{(2\pi)^4}\frac{-2}{(Z^{\dagger}Z)^{3/2}}
\left[(m+r\sum_{\lambda}(c_{\lambda}
-1)c_{\mu}c_{\nu}+2rc_{\mu}s_{\nu}^2\right]a^2F_{\mu\nu}(x)+
O(a^4),
\end{eqnarray}
where $s_{\mu} = sin\,k_{\mu}, c_{\mu} = cos\,k_{\mu}$ and
\begin{equation}
Z^{\dagger}Z = \sum_{\nu}s_{\nu}^2  + \left[ m+
r\sum_{\mu}(c_{\mu} -1)\right]^2.
\end{equation}

 For the case where $r=1, m= 1.368$ (which
corresponds to Wilson $\kappa = 0.19$ in the kernel), the
proportional constant is $-0.08156$ for the overlap
fermion~\cite{lah06}. This has been numerically
verified~\cite{ahl06}. One can try to use this operator to
calculate glue properties, such as the glue momentum and angular
momentum fractions in the nucleon to see if they are better than
the glue operators constructed from the link variables as is in
the case of the local topological charge.

\section{Monte Carlo Simulation}

Take the proportional constant for the classical continuum limit
in Eq. (\ref{action}) to be $c$ which has been
evaluated~\cite{ahl06} as a function of the negative mass
parameter $\rho$ in $D_W$, the kernel of the overlap operator
$D_{ov}$. Then the lattice gauge action can be written as
\begin{equation}
S_g = \frac{1}{2cg^2}{\rm Tr}\, (D_{ov}-D_{ov}^0{}),
\end{equation}
where $Tr$ denotes the trace over all indices, e.g. color, spin,
and space-time. For a given lattice, the trace over the free quark
overlap operator, i.e. $Tr\, D_{ov}^0{}$ is a constant which has
no effect on the Markov process to obtain equilibrium gauge
configurations, we shall drop it. Noting that $D_{ov}$ obeys
$\gamma_5$ hermiticity and satisfies Ginsparg-Wilson relation, we
have
\begin{equation}
S_g = \frac{1}{2cg^2}{\rm Tr}\, D_{ov} = \frac{1}{4cg^2}{\rm
Tr}\,(D_{ov}+ D_{ov}^{\dagger}) = \frac{1}{4cg^2}{\rm Tr}\,
D_{ov}^{\dagger}D_{ov}.
\end{equation}

Now the lattice QCD partition function is
\begin{equation}  \label{partition}
Z = \int {\mathcal{D}}U\,d\bar{\psi}_fd\psi_f e^{-S_g + \sum_{f
=1}^{N_f} \bar{\psi}_f(D_{ov}(m_f))\psi_f},
\end{equation}
where $D_{ov}(m_f)$ is the overlap operator for a quark with mass
$m_f$
\begin{equation}
D_{ov}(m_f) = \rho D_{ov} + m_f (1 - \frac{1}{2}D_{ov}),
\end{equation}
where $\rho$ is the negative mass parameter in $D_W$. Since $e^{Tr
M} = e^{{\rm Tr} \ln e^M} = \det e^M$, the gauge part of the
partition function in Eq. (\ref{partition}) can be written in
terms of a fictitious fermion field $\psi_g$ so that
\begin{equation}
Z = \int {\mathcal{D}}U d\bar{\psi}_gd\psi_g d\bar{\psi}_fd\psi_f
e^{\bar{\psi}_g( e^{-\frac{1}{4cg^2}
D_{ov}^{\dagger}D_{ov}})\psi_g + \sum_{f =1}^{N_f}
\bar{\psi}_f(D_{ov}(m_f))\psi_f}.
\end{equation}

 After integration of the fermion fields, it can be
written as
\begin{equation}
Z = \int {\mathcal{D}}U \det(e^{-\frac{1}{4cg^2}
D_{ov}^{\dagger}D_{ov}})\prod_{f=1}^{N_f} \det(D_{ov}(m_f)).
\end{equation}
We see that the gauge action plays the role of a UV-filtering for
the fermion action (note that $c > 0$ for the range of parameters
for the Wilson kernel in the overlap operator with $r = 1$ and $2
> m > 1$~\cite{ahl06}). The efficiency of an UV-filtered fermion determinant
has been studied by Duncan, Eichten, and Thacker~\cite{det99} and
by Borici~\cite{bor02}.

One way to carry out Monte Carlo simulation is to use
pseudofermions to simulate the determinant. For example, one can
equally split the gauge determinant and attach them to the fermion
determinants of different flavor. In terms of the pseudofermions,
it is
\begin{equation}
Z=\int {\mathcal{D}}U\, d\phi_f^{*}\,d\phi_f
\,e^{-\sum_{f=1}^{N_f}\phi_f^{*}e^{\frac{1}{4cN_fg^2}
D_{ov}^{\dagger}D_{ov}}D_{ov}^{-1}(m_f)\phi_f}.
\end{equation}

We shall discuss two ways to approximate the pseudofermion action.
Since $D_{oc}$ is normal, i.e. $[D_{ov}^{\dagger}, D_{ov}] = 0$,
one can write
\begin{equation}
\phi_f^{*}\,e^{\frac{1}{4cN_fg^2}
D_{ov}^{\dagger}D_{ov}}D_{ov}^{-1}(m_f)\phi_f =
\phi_f^*\,e^{\frac{1}{8cN_fg^2}D_{ov}^{\dagger}D_{ov}}D_{ov}^{-1}(m_f)
e^{\frac{1}{8cN_fg^2}D_{ov}^{\dagger}D_{ov}}\phi_f.
\end{equation}
The range of eigenvalues of $D_{ov}^{\dagger}D_{ov}$ is from 0 to
$4\rho^2$. If $\frac{1}{2cN_fg^2}$ is about unity or less, one can
consider the Chebyshev polynomial approximation to degree $M$
\begin{equation}
e^{\frac{1}{8cN_fg^2}D_{ov}^{\dagger}D_{ov}} \sim \sum_{i=1}^M c_i
(D_{ov}^{\dagger}D_{ov})^i.
\end{equation}
Alternatively, one can perform a Chebyshev rational polynomial
approximation for the operator $e^{\frac{1}{4cN_fg^2}
D_{ov}^{\dagger}D_{ov}}D_{ov}^{-1}(m_f)$ to degree $N$ for the
Rational Hybrid Monte Carlo algorithm (RHMC)~\cite{hks99}. For the
case of $2+1$ flavors, the pseudofermion action for the 2
degenerate flavors can be approximated by
\begin{equation}   \label{pf2}
\phi^*\,e^{\frac{1}{6cg^2}D_{ov}^{\dagger}D_{ov}}(D_{ov}^{\dagger}
D_{ov}(m_f))^{-1}\phi \sim \phi^* \sum_{i=1}^N
\frac{a_i}{D_{ov}^{\dagger}D_{ov} + b_i}\phi,
\end{equation}
and the single flavor one approximated by
\begin{equation}  \label{pf1}
\phi^*\,e^{\frac{1}{12cg^2}D_{ov}^{\dagger}D_{ov}}(D_{ov}^{\dagger}
D_{ov}(m_f))^{-1/2}\phi \sim \phi^* \sum_{i=1}^N
\frac{c_i}{D_{ov}^{\dagger}D_{ov} + d_i}\phi.
\end{equation}

  In this case, the forces in the equation of motion in HMC come
from the effective pseudofermion actions in Eqs. (\ref{pf2}) and
(\ref{pf1}) which represent the combined gauge  and fermion
forces.

     If one uses a multi-mass algorithm for inversion and the coefficients
$b_i$ and $d_i$ are not smaller than $m_f^2$, the overhead of
incorporating the gauge action in RHMC is negligible compared to
the ordinary $2+1$ flavor simulation in HMC with the same
inverter.

One can of course accelerate the Rational Hybrid Monte Carlo
algorithm (RHMC) by splitting the determinant into fractional
flavors~\cite{ck04,jhl03} with more pseudofermion fields and
improve the overall efficiency as shown by Clark and
Kennedy~\cite{ck04}.

In summary, we have discussed possible Monte Carlo simulations of
the lattice gauge action from the trace of the overlap operator,
i.e. ${\rm Tr}\, D_{ov}$ together with the overlap fermion action
in the context of HMC. By virtue of the fact that the overlap
operator is exponentially local, the gauge action so defined is
expected to behave like a chirally smeared action. Furthermore,
the integrand of the gauge part of the partition function, written
in terms of a determinant, appears to be an UV-filter for the
fermion determinant. We should note that, similar to the overlap
fermion action, this gauge action is not reflection positive. Also
presented is our derived result of the lattice gauge field tensor
as the classical continuum limit of ${\rm tr_s}\,\sigma_{\mu\nu}
D_{ov}(x,x)$. This can be used to calculate glue matrix elements
in the hadrons and possibly glueballs.

This work is partially supported by DOE grant DE-FG05-84ER40154.
We wish to thank I.~Horv\'{a}th, A. Alexandur and A. Kennedy for
stimulating discussions. The author also wish to thank the
hospitality of the Institute of Physics, Academia Sinica, Taipei,
Taiwan, National e-Science Center, Edinburgh and DESY-Zeuthen,
Germany where the talk is prepared and the proceedings is written
up during the visit of these places.


\end{document}